\documentclass{PoS}
\usepackage{amsmath,amssymb,graphicx}

\newcommand{\Nt}{N_\tau}
\newcommand{\muB}{\mu_B}
\newcommand{\muQ}{\mu_Q}
\newcommand{\muS}{\mu_S}
\newcommand{\veps}{\varepsilon}
\newcommand{\bmu}[1]{\left(\frac{\mu_B}{T}\right)^#1}
\newcommand{\ord}{\mathcal{O}}
\newcommand{\sNN}{s_{NN}^{1/2}}

\title{The QCD Equation of State to $\ord(\mu_B^4)$}

\ShortTitle{EoS to $\ord(\mu_B^4)$}

\author{\speaker{Prasad Hegde} (for the BNL-Bielefeld-CCNU collaboration)\thanks{The members of the BNL-Bielefeld-CCNU collaboration are: A.~Bazavov, H.-T.~Ding, P.~Hegde, O.~Kaczmarek, F.~Karsch, E.~Laermann, Y.~Maezawa, S.~Mukherjee, H.~Ohno, P.~Petreczky, C.~Schmidt, S.~Sharma, W.~S\"oldner and M.~Wagner.}\\
        Key Laboratory of Quark \& Lepton Physics (MOE) \\
        and Institute of Particle Physics, \\
        Central China Normal University,\\
        Wuhan 430079, China.\\
        E-mail: \email{phegde@mail.ccnu.edu.cn}}

\abstract{We present results from an ongoing calculation of the QCD equation of state at finite baryon chemical potential $\muB$. We use the method of Taylor expansions to circumvent the sign problem and calculate the expansion coefficients to sixth order using HISQ fermions~\cite{Follana:2006rc}. We work at two lattice spacings, namely $\Nt=6$ and 8 and, though we do not take the continuum limit, demonstrate that cutoff effects remain under control. We also use our results to construct an equation of state along the freeze-out curve. Using our sixth-order results as a cross-check, we demonstrate that our fourth-order equation of state is suitable for the modeling of dense matter created in heavy ion collisions with center-of-mass energies down to $\sNN\sim20$ GeV.}

\FullConference{The 32nd International Symposium on Lattice Field Theory,\\
		23-28 June, 2014\\
		Columbia University New York, NY}

\begin{document}
\section{Introduction}
\label{sec:intro}
The QCD equation of state (EoS) is a crucial input in modeling the hydrodynamic evolution of thermal matter created in heavy-ion collisions. It is well-known that such matter is strongly-interacting both at the RHIC and at the LHC and therefore calculating the EoS in this domain requires a non-perturbative first-principles approach such as lattice QCD.

Lattice calculations of the QCD EoS have grown more refined with time. The current state-of-the-art is continuum-extrapolated results for the equation of state at zero baryon chemical potential calculated using staggered fermions~\cite{Bazavov:2014pvz, Borsanyi:2013bia}. With the advent of the Beam Energy Scan (BES) experiment at the Relativistic Heavy Ion Collider (RHIC) however, it has become necessary to extend these results to moderately large values of the baryon chemical potential $\muB$\footnote{Specifically, one expects to reach $\muB\approx$ 400-450 MeV at the lowest center-of-mass energies~\cite{Cleymans:1999st}.}. As is well-known, such an extension is non-trivial because of the sign problem of lattice QCD~\cite{deForcrand:2010ys}. While a complete solution to this problem is not yet known, various partial solutions exist~\cite{Allton:2002zi, Gavai:2001fr, Fodor:2002km, D'Elia:2002gd, Ejiri:2005uv, Aarts:2009hn, Cristoforetti:2012su}. Among these, the method of Taylor expansions is the most straightforward~\cite{Gavai:2001fr, Allton:2002zi}. It has the advantage that the uncertainty coming from the truncation of the Taylor series can be straightforwardly estimated from a knowledge of the next coefficient. Moreover, the various coefficients of the expansion bear a straightforward interpretation, either as cumulants (diagonal coefficients) or as correlations (off-diagonals) between conserved charges. Because of this, they can be used to probe deconfinement~\cite{Bazavov:2012jq}, explore the degrees of freedom at a given temperature~\cite{Bazavov:2013dta,Bazavov:2013uja} or determine freeze-out conditions~\cite{Bazavov:2012vg}. They can also be determined experimentally from the moments of various hadron multiplicity distributions~\cite{Adamczyk:2014fia}.

The starting point of this method is the expansion of the partition function in powers of the chemical potentials, namely
\begin{equation}
\frac{p}{T^4} = \frac{1}{VT^3}\ln\mathcal{Z}(\mu_u,\mu_d,\mu_s) = \sum_{i,j,k=0}^\infty%
\frac{\chi_{ijk}}{i!j!\,k!}%
\left(\frac{\mu_B}{T}\right)^i \left(\frac{\mu_Q}{T}\right)^j \left(\frac{\mu_S}{T}\right)^k
\longrightarrow
\sum_{n=0}^\infty c_n\left(\frac{\muB}{T}\right)^n.
\label{eq:definition}
\end{equation}

With three flavors of quarks, one has three chemical potentials. A change of basis allows us to express these in terms of conserved charge chemical potentials, namely baryon number, electric charge and strangeness $(\muB,\muQ,\muS)$. Of these, $\muQ$ and $\muS$ may be determined as functions of $\muB$ by imposing the constraints coming from the initial conditions in heavy-ion collisions namely, zero net strangeness ($n_S=0$) and a fixed $Z$-to-$A$ ratio ($n_Q/n_B=r$)~\cite{Bazavov:2012vg}
\begin{align}
& & \frac{\muQ}{T} = q_1\frac{\muB}{T} + q_3\left(\frac{\muB}{T}\right)^3+\dots,
& & \frac{\muS}{T} = s_1\frac{\muB}{T} + s_3\left(\frac{\muB}{T}\right)^3+\dots. & &
\end{align}

\begin{figure}[!htb]
\hspace{-0.035\textwidth}%
\includegraphics[width=0.37\textwidth]{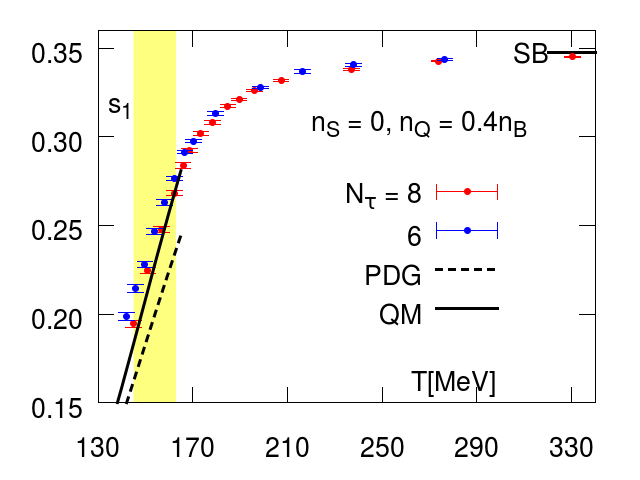}%
\hspace{-0.02\textwidth}%
\includegraphics[width=0.36\textwidth]{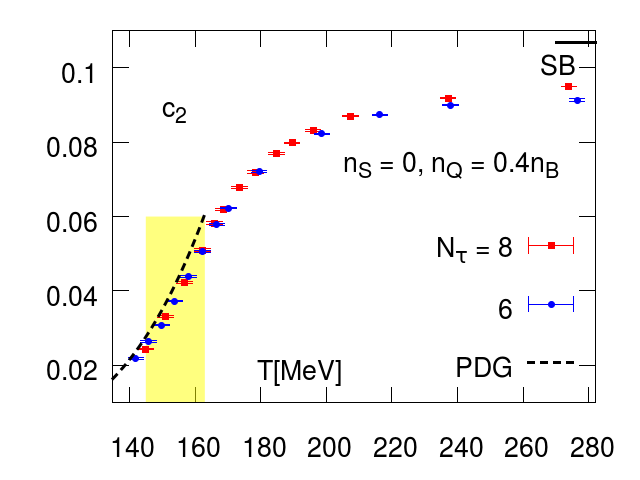}%
%\hspace{0.02\textwidth}%
\includegraphics[width=0.36\textwidth]{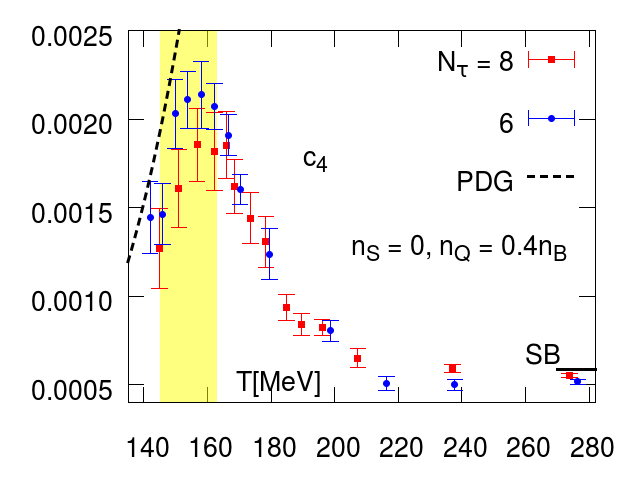}%
\caption{(Left) The first Taylor coefficient in the expansion of $\muS/T$. The dashed black line is from an HRG model containing all the Particle Data Group resonances (henceforth PDG-HRG) while the solid black line also takes into account additional light and strange resonances coming from the Quark Model~\cite{Bazavov:2014xya} (henceforth QM-HRG). (Center and Right) The first two Taylor coefficients in the expansion of  $p/T^4$. In all three figures, the shaded yellow region indicates the physical quark mass chiral crossover region, $T_c=154(9)$ MeV~\cite{Bazavov:2011nk}. $r=0.4$ is the value for the $Z$-to-$A$ ratio in Pb-Pb collisions.}
\label{fig:cn}
\end{figure}

We plot $s_1$ for our two lattice spacings in Fig.~\ref{fig:cn}. Its mild cutoff dependence and small errors enabled us to provide an estimate for the continuum limit in Ref.~\cite{Bazavov:2014xya}. This estimate was found to exceed the predictions of the usual Hadron Resonance Gas (HRG) model below the chiral crossover temperature. We found that this difference could be accounted for by including in our HRG model additional, as yet unobserved light and strange resonances which are predicted by both lattice QCD and the Quark Model~\cite{Bazavov:2014xya, Edwards:2012fx, Capstick:1986bm, Ebert:2009ub}. As regards the other coefficients, the coefficient $q_1$ is negative and only about 1\% of $s_1$ for the whole temperature range that we consider. Similarly, both $s_3/s_1$ and $q_3/q_1$ are only around 10\% in magnitude at $T\approx150$ MeV and not more than 1-2\% for $T\gtrsim170$ MeV~\cite{Bazavov:2012vg}.

Fig.~\ref{fig:cn} also shows our results for the first two Taylor coefficients $c_2$ and $c_4$.\footnote{All odd coefficients are zero at $\muB=0$ by $CP$ symmetry.} It is seen that the signal-to-noise ratio drops quickly as we go to higher orders. This is actually due to the sign problem, which manifests as a noise problem at $\muB=0$~\cite{Endres:2011jm}. To overcome this, we generated around 100,000-200,000 trajectories for the temperatures near the crossover region. We also used up to 1,500 random vectors per configuration to determine the operator traces that were required to calculate the susceptibilities\footnote{A list of all the traces that are required for the sixth-order calculation is given in~\cite{Allton:2005gk}.}.

\section{First-Order Quantities}
\label{sec:first_order}
From the pressure, the conserved charge, entropy and energy densities can be obtained by taking derivatives with respect to the temperature
\begin{align}
\frac{n_i}{T^3} =T\frac{\partial}{\partial\mu_i}\left(\frac{p}{T^4}\right),  &&
\frac{s}  {T^3} = \left[T\frac{\partial}{\partial T}-4\right]\frac{p}{T^4}, &&
\frac{\veps}{T^4}     = \frac{Ts+\sum_i\mu_in_i-p}{T^4}.
\label{eq:thermo}
\end{align}
We get the order-by-order corrections by rewriting these in terms of the $c_n$'s\footnote{The number densities are not directly expressible in terms of the $c_n$'s but must be calculated and fitted separately.}% However, note that $\sum_i\mu_in_i/T^4 = \sum_{n=0}^\infty \bmu{n}\cdot nc_n$.}
\begin{align} && &&
\frac{\veps}{T^4} &= \sum_{n=0}^\infty \bmu{n} \left\{T\frac{dc_n}{dT} +   3  c_n\right\},&& &&
\frac{s}{T^3}     &= \sum_{n=0}^\infty \bmu{n} \left\{T\frac{dc_n}{dT} + (4-n)c_n\right\}.&& &&
%\label{eq:thermo}
\end{align}
We calculated these derivatives in the temperature range 145 MeV $\leqslant T \leqslant$ 220 MeV from spline fits to the $c_n$. We divided the temperature range into intervals 145 MeV $<T_1<\dots<$ 220 MeV by choosing breakpoints $T_i$, and sought to obtain the best possible fit by varying both the number and the locations of the breakpoints. We used quartic splines instead of the more commonly used cubic splines because we found that they interpolated the data better and with fewer breakpoints. They also had the advantage that their second derivatives were smooth rather than merely continuous. Unlike the $\muB=0$ case in which the location of the breakpoints too was determined by the fitting routine~\cite{Bazavov:2014pvz}, here we chose to vary the breakpoints by hand. We could do this because only a small number of breakpoints was necessary to get good results, due to both the narrow temperature range as well as the fact that we fitted the $\Nt=6$ and 8 datasets separately.

\begin{figure}[!htb]
\includegraphics[width=0.45\textwidth]{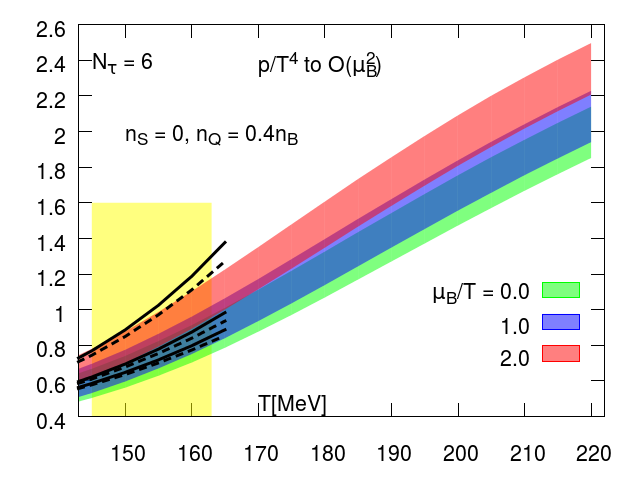}%
\hspace{0.10\textwidth}%
\includegraphics[width=0.45\textwidth]{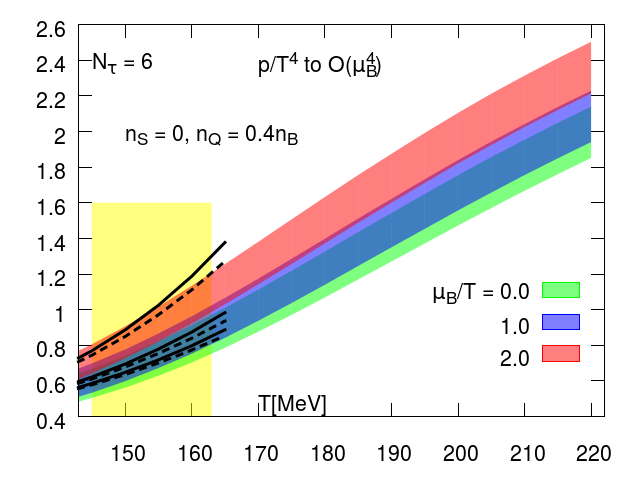}
\includegraphics[width=0.45\textwidth]{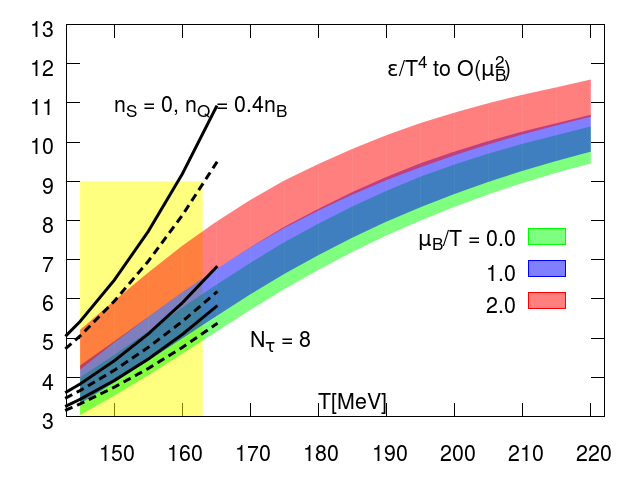}%
\hspace{0.10\textwidth}%
\includegraphics[width=0.45\textwidth]{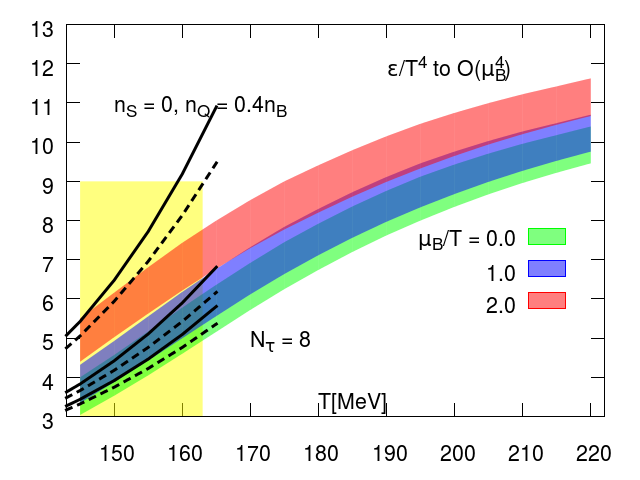}
\caption{(Above) Total $p/T^4$ and (Below) total $\veps/T^4$, to $\ord(\muB^2)$ and $\ord(\muB^4)$, for different values of $\muB/T$. These figures are all for the constrained case with $r=0.4$. The dashed and solid black lines are PDG-HRG and QM-HRG curves respectively.}
\label{fig:total}
\end{figure}

Once we had finalized the fits, we could add the contributions from different orders to the $\muB=0$ results one at a time to obtain the $\ord(\mu_B^2)$, $\ord(\mu_B^4)$ and $\ord(\mu_B^6)$ approximations respectively for a given value of $\muB/T$. However since our sixth-order susceptibilities were noisy, we will restrict ourselves to a fourth-order equation of state from now on and use the sixth-order results just as a cross-check. 

\begin{figure}[!htb]
\includegraphics[width=0.45\textwidth]{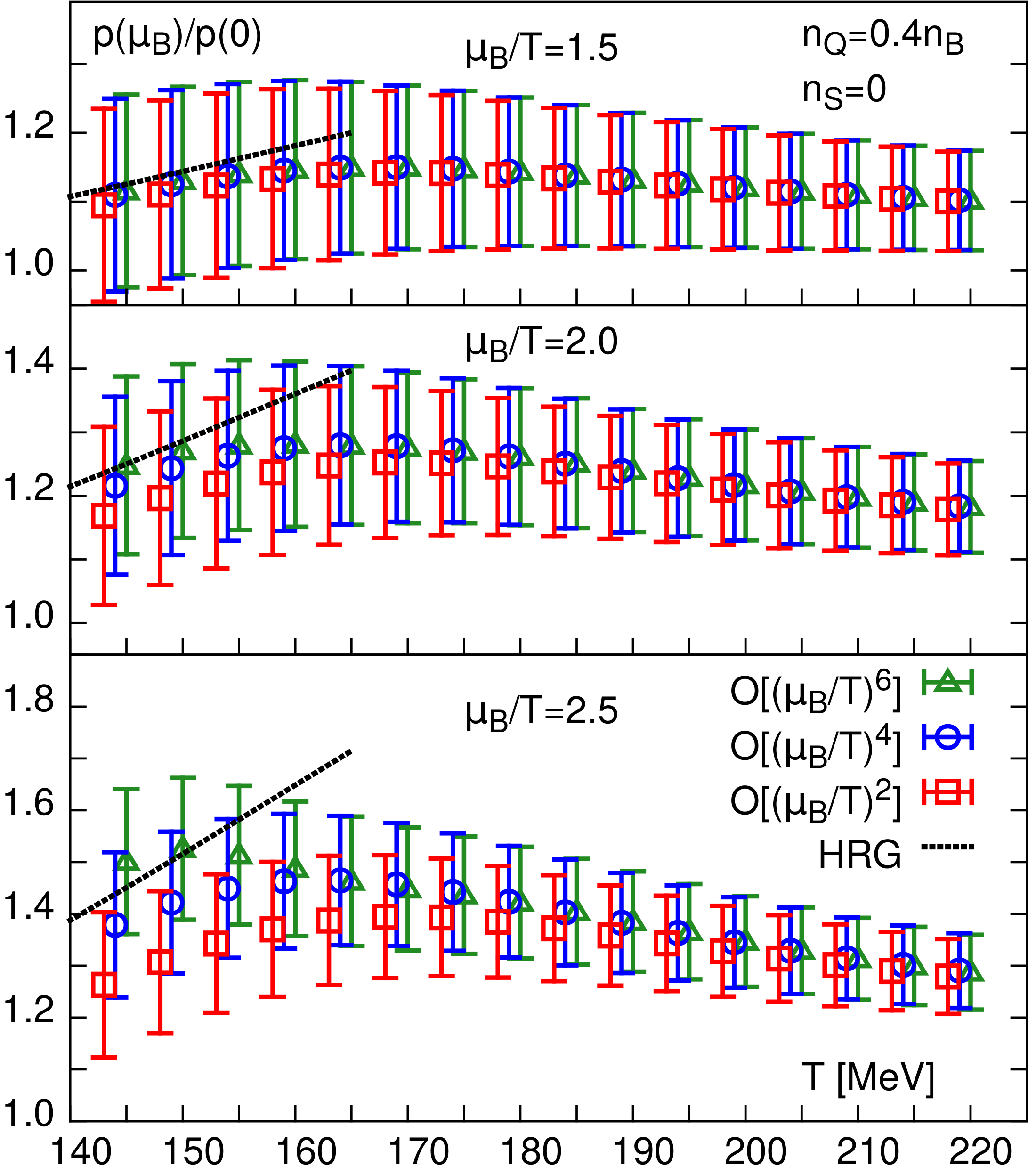}%
\hspace{0.10\textwidth}%
\includegraphics[width=0.45\textwidth]{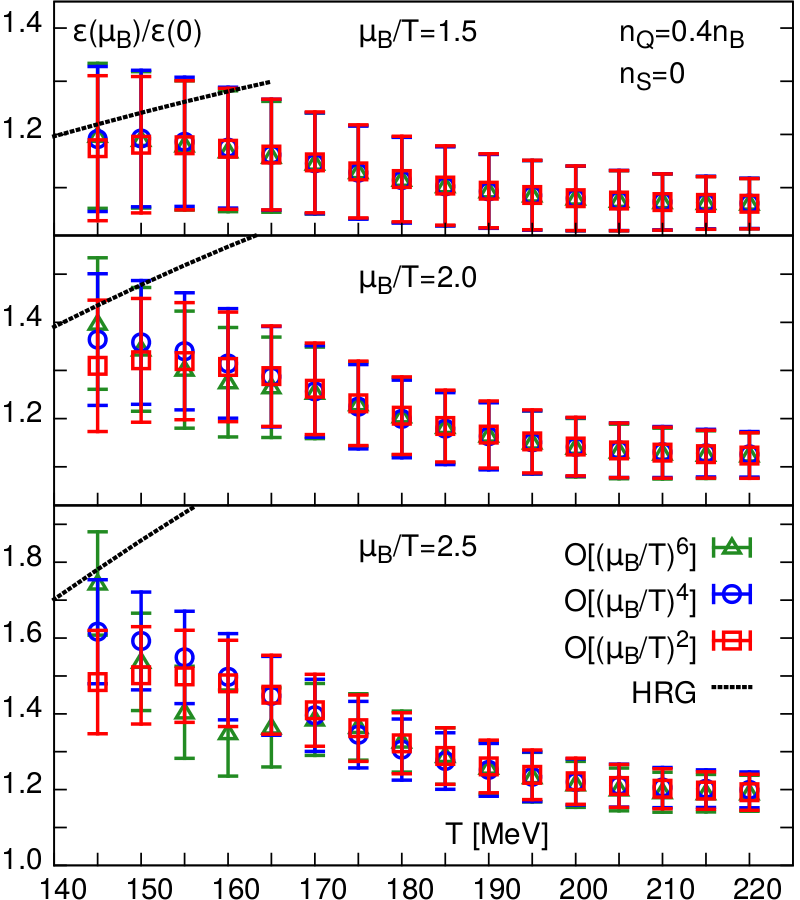}%
\caption{(Left)The ratio $\Delta (p/T^4)/(p/T^4)_{\muB=0}$ where $\Delta$ represents the finite $\muB$ contribution upto second, fourth and sixth orders for the constrained case with $r=0.4$. (Right)The same for $\veps/T^4$. The dashed black lines are PDG-HRG lines.}
\label{fig:ratios}
\end{figure}

Fig.~\ref{fig:total} presents our results for the pressure and energy. One sees that the dominant corrections come from the second-order term for small values of $\muB/T$, just as one might expect. Higher-order corrections start to become significant beyond $\muB/T\gtrsim2.0$. This is also seen from Fig.~\ref{fig:ratios}, where we plot the ratio of second-, fourth- and sixth-order pressure and energy to the zeroth-order result. For $\muB/T\simeq2.0$ and beyond, the ratio starts to depend on the order at which we truncate the expansion.

\section{Observables on the Freezeout Curve}
\label{sec:freezeout}
\begin{figure}[!thb]
\centering
\includegraphics[width=0.4\textwidth]{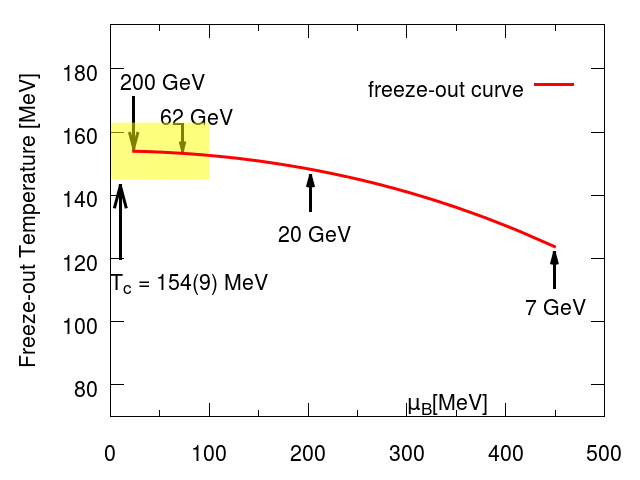}
\caption{The freeze-out temperature and chemical potential as the beam energy is decreased. The vertical extent of the shaded yellow rectangle is the chiral crossover region for physical quark masses.\label{fig:fzTmu}}
\end{figure}

From a statistical model fit to hadron multiplicities at freeze-out, one can extract the temperature $T^f$ and baryon chemical potential $\muB^f$ at freeze-out~\cite{BraunMunzinger:2003zd}. Moreover, one can parametrize $(T^f,\muB^f)$ as functions of the beam energy $\sNN$ as (with $x=\sNN$ in GeV)~\cite{Cleymans:1999st}
\begin{align}
& & \frac{T^f}{T^f_\infty} = \frac{1}{1+\exp\left(1.176-\frac{\ln x}{0.45}\right)},
& & \mu_B^f = \frac{1.303}{1+0.286\,x} \; \text{GeV}. & &
\end{align}

We use $T^f_\infty=154$ MeV, in accordance with lattice results for the crossover temperature~\cite{Bazavov:2011nk}. Fig.~\ref{fig:fzTmu} shows $T^f$ and $\muB^f$ for a few of the beam energies from the RHIC BES program. By plugging the values obtained from this parametrization into our expansions, we can determine the values of various thermodynamic observables at freeze-out as a function of the beam energy. As mentioned earlier, such an ``equation of state along the freeze-out curve'' should be useful in the context of the Beam Energy Scan program.

\begin{figure}[!htb]
\hspace{-0.045\textwidth}
\includegraphics[width=0.37\textwidth]{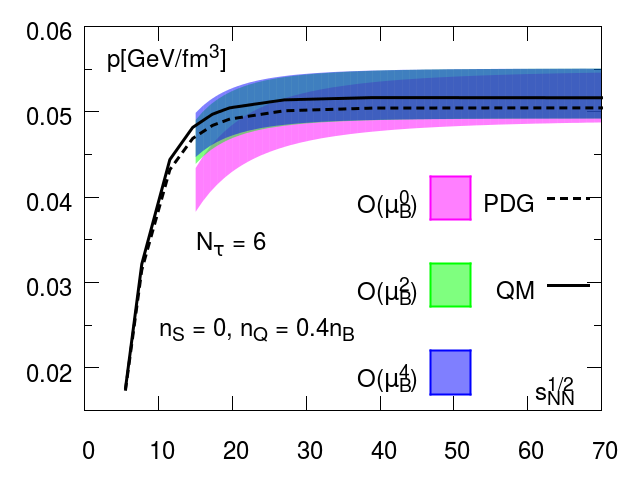}%
\hspace{-0.02\textwidth}%
\includegraphics[width=0.36\textwidth]{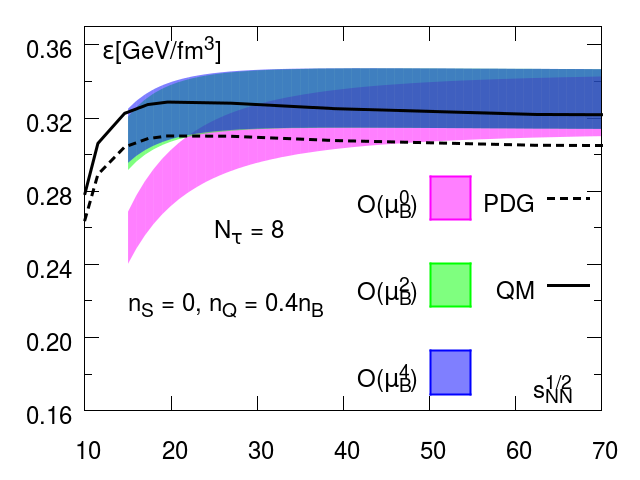}%
%\hspace{-0.005\textwidth}
\includegraphics[width=0.36\textwidth]{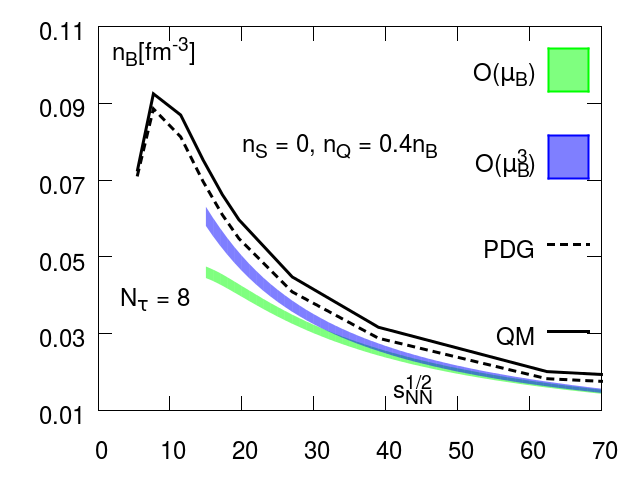}
\caption{Pressure, energy density and baryon number density respectively on the freeze-out curve. Also shown are the predictions from PDG and QM Hadron Resonance Gas models. Note that once $\ord(\muB^4)$ corrections are included, both $p$ and $\veps$ remain constant along the freeze-out curve down to $\sNN\sim20$ GeV.}
\label{fig:freezeout}
\end{figure}

We present our results for the pressure, energy density and baryon density in Fig.~\ref{fig:freezeout}, along with the PDG-HRG and QM-HRG predictions. We see that both $p$ and $\veps$ are constant along the freeze-out curve for all except the lowest beam energies. In part, this is because the largest contribution to both these quantities comes from the zeroth order, as a result of which they are more sensitive to changes in $T^f$ than $\muB^f$. Note however that it is necessary to include the $\ord(\muB^4)$ contribution for the curves to remain flat down to $\sNN\sim30$ GeV. Below this energy, finite-$\muB$ corrections start to become comparable to the zeroth-order values. Similarly, it is necessary to take the $\ord(\muB^3)$ corrections into account for $n_B$ below this beam energy. We expect that sixth-order corrections will also have to be taken into account below $\sNN\sim20$ GeV, but better results for the sixth-order susceptibilities are needed to determine the exact value.

\section{Conclusions}
We have attempted to extrapolate the QCD equation of state for $\muB=0$ to nonzero values of $\muB$ through the method of Taylor expansions. We found that by carrying out the expansion to sixth order, we could obtain reliable results for the pressure and energy density up to $\muB/T\sim2.0$ for temperatures at and above the crossover. Since we still need better statistics for our sixth-order results, we merely used them as a cross-check in constructing a fourth-order equation of state. Lastly, by plugging the values of $(T^f,\muB^f)$ obtained from a parametrization to the freeze-out curve, we found that this fourth-order EoS was useful down to beam energies $\sNN\sim20$ GeV.

\section*{Acknowledgements}
The author is partially supported by grant QLPL2014P01 of the Ministry of Education of the People's Republic of China. The numerical calculations described here have been performed at JLab and at Indiana University in the United States and at Bielefeld University and Paderborn University in Germany. We acknowledge the support of Nvidia through the CUDA Research Center at Bielefeld University.

\end{document}